\documentclass{article}
\usepackage{preprint,times}

\usepackage{amsmath,amsfonts,bm}

\def\eqref#1{equation~\ref{#1}}

\def\1{\bm{1}}

\DeclareMathAlphabet{\mathsfit}{\encodingdefault}{\sfdefault}{m}{sl}
\SetMathAlphabet{\mathsfit}{bold}{\encodingdefault}{\sfdefault}{bx}{n}

\usepackage[colorlinks=true,linkcolor=blue,citecolor=blue,urlcolor=blue]{hyperref}
\usepackage{url}
\usepackage{graphicx}
\usepackage{subcaption}
\usepackage{amsthm}
\usepackage{amsmath}
\usepackage{amssymb}
\usepackage{mathtools}
\usepackage{pifont}
\usepackage{booktabs}
\usepackage{arydshln}
\usepackage{xspace}
\usepackage{algorithm}
\usepackage{algorithmicx}
\usepackage{algpseudocode}
\usepackage{multirow}
\usepackage{wrapfig}

\makeatletter
\DeclareRobustCommand\onedot{\futurelet\@let@token\@onedot}
\def\@onedot{\ifx\@let@token.\else.\null\fi\xspace}

\def\eg{\emph{e.g}\onedot} 
\def\ie{\emph{i.e}\onedot}

\DeclareRobustCommand{\ourapproach}{{\sc CodeRL+}\xspace}
\DeclareRobustCommand{\ourapproachbf}{{\sc \textbf{CodeRL+}}\xspace}
\makeatother

\theoremstyle{definition}

\theoremstyle{remark}

\usepackage{tcolorbox}
\tcbuselibrary{breakable}
\usepackage{xcolor}

\newtcolorbox[use counter=prompt]{promptbox}[2][]{%
  colback=gray!5,
  colframe=black!50,
  fonttitle=\bfseries,
  title=Prompt: #2,
  breakable,
  #1
}

\iclrfinalcopy
\title{ 
\ourapproach: Improving Code Generation via Reinforcement with Execution Semantics Alignment
}

\author{Xue Jiang$^{1,2}$, Yihong Dong$^{1,2}$, Mengyang Liu$^{1}$, Hongyi Deng$^{1}$, Tian Wang$^{1}$, Yongding Tao$^{1}$, \\ \textbf{Rongyu Cao$^{2}$, Binhua Li$^{2}$, Zhi Jin$^{1,3}$, Wenpin Jiao$^{1}$, Fei Huang$^{2}$, Yongbin Li$^{2}$, Ge Li$^{1}$}\\
$^1$ School of Computer Science, Peking University \\$^2$ Tongyi Lab, Alibaba Group\\$^3$ School of Computer Science, Wuhan University\\
\texttt{\{jiangxue, dongyh\}@stu.pku.edu.cn} \quad \texttt{lige@pku.edu.cn}\\\
}

\begin{document}

\maketitle

\begin{abstract}
While Large Language Models (LLMs) excel at code generation by learning from vast code corpora, a fundamental semantic gap remains between their training on textual patterns and the goal of functional correctness, which is governed by formal execution semantics. Reinforcement Learning with Verifiable Rewards (RLVR) approaches attempt to bridge this gap using outcome rewards from executing test cases. However, solely relying on binary pass/fail signals is inefficient for establishing a well-aligned connection between the textual representation of code and its execution semantics, especially for subtle logical errors within the code.
In this paper, we propose \ourapproach, a novel approach that integrates execution semantics alignment into the RLVR training pipeline for code generation. \ourapproach enables the model to infer variable-level execution trajectory, providing a direct learning signal of execution semantics. \ourapproach can construct execution semantics alignment directly using existing on-policy rollouts and integrates seamlessly with various RL algorithms.
Extensive experiments demonstrate that \ourapproach outperforms post-training baselines (including RLVR and Distillation), achieving a 4.6\% average relative improvement in pass@1. \ourapproach generalizes effectively to other coding tasks, yielding 15.5\% and 4.4\% higher accuracy on code-reasoning and test-output-generation benchmarks, respectively. \ourapproach shows strong applicability across diverse RL algorithms and LLMs. Furthermore, probe analyses provide compelling evidence that \ourapproach strengthens the alignment between code's textual representations and its underlying execution semantics.
\end{abstract}

\renewcommand{\thefootnote}{\fnsymbol{footnote}}

\footnotetext[1]{Work done during Xue Jiang and Yihong Dong's internship at Tongyi Lab.}
\footnotetext[2]{Our source code is released at \url{https://github.com/jiangxxxue/CODERLPLUS}.}

\renewcommand{\thefootnote}{\arabic{footnote}}

\section{Introduction}
Code generation has become a fundamental capability of Large Language Models (LLMs) and serves as a critical benchmark for evaluating their reasoning and problem-solving abilities~\citep{guo2025deepseek, gemini2.5, gpt4}. From solving complex algorithmic problems~\citep{alphacode,codereval} to developing software projects autonomously~\citep{Self-Collaboration,Self-Planning,classgen,Agent4code}, LLMs are progressively reshaping modern development practices through their code generation capabilities. When evaluating the code generation performance of LLMs, functional correctness stands as the paramount criterion~\citep{function_style,chatgptcode,codereval}, \ie, whether the generated code produces the expected outputs for given inputs. Functional correctness is determined by the code's execution semantics, which are defined by a set of formal, deterministic rules that specify how each statement transforms program state and determines the code's actual behavior~\citep{jain2024livecodebench}.

The fundamental challenge in code generation lies in the semantic gap between the textual representation of LLMs and execution semantics. LLMs acquire their foundational code generation abilities through self-supervised pre-training on code corpora. 
This learning approach trains models to capture the textual patterns of code through autoregressive next-token prediction. However, the correctness of code is not determined by its textual form, but by its execution semantics. Since LLMs receive no direct supervision from functional tests or execution outcomes during pre-training, a fundamental misalignment exists between the LLM's pre-training objective (fitting textual distributions) and the final evaluation criterion (correct execution). Current post-training approaches employ Reinforcement Learning with 
Verifiable Rewards (RLVR) to bridge this semantic gap~\citep{tester_coder,RLVR-limit}. 
RLVR exploits the verifiability of code, where generated solutions can be executed against test cases to provide deterministic feedback, enabling models to optimize directly for functional correctness.

\begin{figure*}[t!]
    \centering
    \begin{subfigure}[b]{0.74\textwidth}
        \centering
        \includegraphics[width=1.02\textwidth]{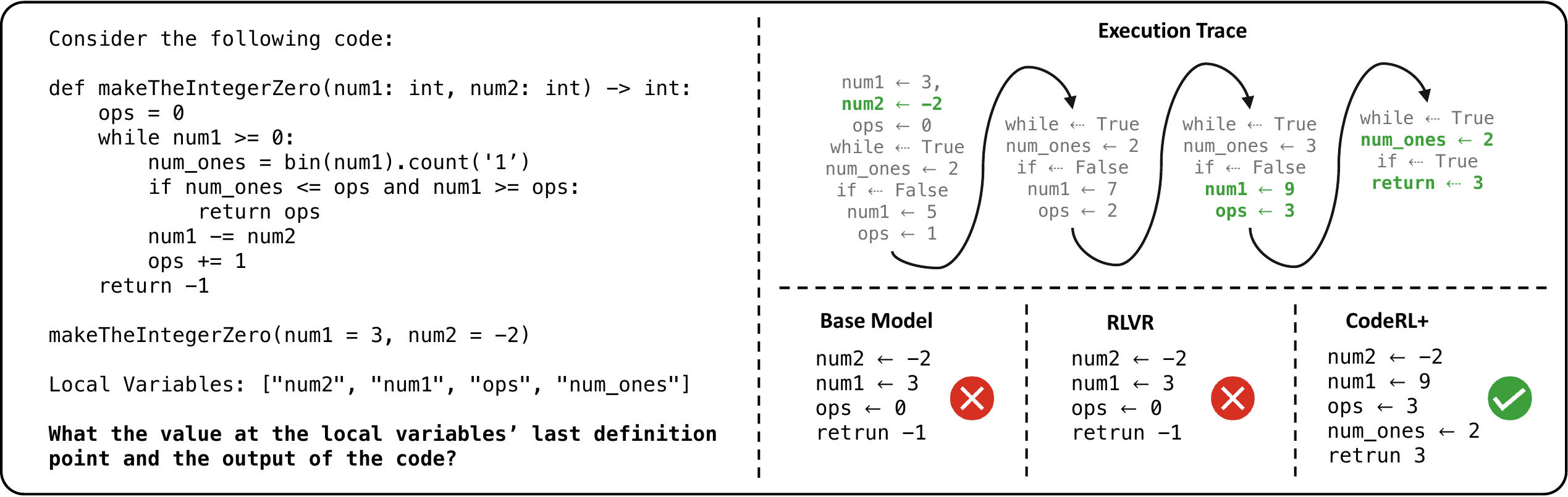}
        \caption{}
        \label{fig:motivation2}
    \end{subfigure}
    \hfill
    \begin{subfigure}[b]{0.24\textwidth}
        \centering
        \includegraphics[width=1.01\textwidth]{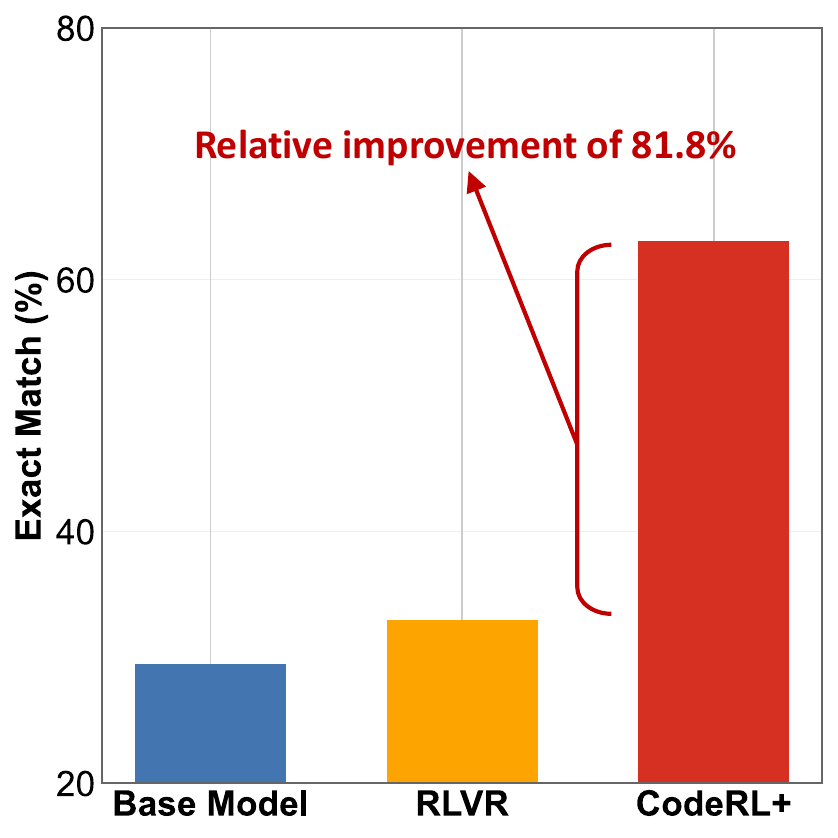}
        \caption{}
        \label{fig:motivation1}
    \end{subfigure}
    \caption{Illustrations of the existing RLVR struggling to establish a well-aligned connection between the textual representation of code and its execution semantics. (a) An example of execution trace inference. (b) Result of execution trace inference task.
    }
    \label{fig:motivation}
\end{figure*}

Despite RLVR's attempts to incorporate execution feedback, empirical evidence reveals that it fails to effectively bridge the semantic gap, which fundamentally limits code generation performance gains. Figure \ref{fig:motivation1} presents results from an execution trace inference task, where models have to infer the final value of each variable in the execution trace. RLVR-trained models show only marginal improvement over base models (4\% increase), indicating that relying solely on sparse pass/fail rewards from final execution outcomes is insufficient. Figure \ref{fig:motivation2} illustrates this concretely: both base and RLVR-trained models fail catastrophically on a loop program, unable to track variable changes through iterations. 
Models that cannot reason about such loop semantics inevitably produce flawed iterative code. This limitation directly impairs code generation performance, often leading to subtle yet critical logical errors. These findings motivate our approach to establishing a stronger, more explicit connection between code's textual representation and its execution semantics in RLVR, rather than depending solely on final execution outcomes.

In this paper, we propose \ourapproach, which advances standard RLVR training for code generation by incorporating execution semantics alignment. Our approach performs parallel reinforcement learning that jointly optimizes code generation and execution semantics alignment, where the latter repurposes failed exploration programs to analyze their underlying execution semantics by inferring how variables propagate during program execution. This integration explicitly aligns the generated code's textual form with its functional behavior, providing dense learning signals that effectively bridge the gap between textual fluency and execution correctness in LLMs. Crucially, execution semantics alignment employs an on-the-fly training scheme that can be dynamically constructed from code generation rollout programs, requiring no additional data source while evolving with the model's capabilities.

Extensive experiments show that \ourapproach achieves state-of-the-art performance over GRPO and recently proposed post-training models and methods on mainstream code generation benchmarks, such as HumanEval, LeetCode, and LiveCodeBench. On more generalized code-related tasks, \ie, reasoning and test output generation tasks, \ourapproach also significantly outperforms baselines, including the methods solely optimized for code reasoning. Additionally, more extensive experiments demonstrate that \ourapproach has stable and consistent improvements across different families, different sizes of language models, and different RLVR algorithms, showcasing the method's strong applicability. A probing experiment proves that after training with \ourapproach, LLMs consider execution semantics more when generating code.

\section{Related Work}
In this section, we outline the two most relevant directions and associated papers of this work.

\subsection{Reinforcement Learning for Code Generation}
Reinforcement learning (RL) has emerged as a potential approach for optimizing code generation beyond pre-training and supervised fine-tuning, which often produce syntactically plausible but functionally incorrect code~\citep{codellama,memorize_or_generalize,ROCODE}\nocite{CodeScore,SEED}. 
Early explorations such as CodeRL~\citep{le2022coderl} employ actor-critic frameworks to leverage unit test feedback for code generation. StepCoder~\citep{dou2024stepcoder} introduces curriculum learning with RL to decompose complex tasks into manageable subtasks, while CodePRM~\citep{li2025codeprm} addresses the sparse reward problem through process reward models that provide dense feedback. The scope of RL applications further expands with RLCoder~\citep{wang2024rlcoder}, which applies RL to learn retrieval strategies for project code completion.
While these works laid the groundwork, their modest performance gains failed to establish RL as a compelling alternative to supervised approaches. However, the landscape shifted with DeepSeek-R1~\citep{guo2025deepseek}, which demonstrated that combining efficient RL algorithms like GRPO with chain-of-thought reasoning can boost problem-solving capabilities of LLMs, reigniting interest in RL for code generation. More recent efforts include jointly optimizing code and unit test generation~\citep{tester_coder}, using RL for adapting to API updates~\citep{wu2025recode}, and rewarding intermediate reasoning steps conditional on correct final outputs~\citep{fan2025posterior}. 

We design \ourapproach from an orthogonal perspective that introduces execution semantics, which can be combined with these RL methods to enhance code generation.

\subsection{Learning Program Executions with Large Language Models} 
Prior work on learning program executions to enhance LLMs' code reasoning capabilities has predominantly employed knowledge distillation from stronger teacher models combined with supervised fine-tuning~\citep{le2025visualcoder,ding2024semcoder,licodeio,cwm2025}. A representative work, CODEI/O~\citep{licodeio}, enhances code reasoning by distilling from DeepSeek-V2.5 and fine-tuning models to predict execution inputs/outputs given code and outputs/inputs. However, distillation methods are inherently bound by the teacher model's capabilities~\citep{xu2024survey,minillm}. Moreover, supervised fine-tuning has been criticized for merely imitating surface patterns rather than genuinely learning reasoning processes~\citep{gudibande2023false,turpin2023}, often resulting in degraded generalization performance on other code-related tasks~\citep{codellama}.
Following the trend of Deepseek-R1 using RL to drive the general reasoning, RLVR pipelines have been applied to code reasoning. CodeReasoner~\citep{codereasoner} is designed to improve LLM code reasoning performance through a two-stage training process combining instruction fine-tuning and GRPO. CodeBoost~\citep{wang2025codeboost} leverages RL solely with code reasoning tasks to address the challenge that collecting high-quality coding instructions for fine-tuning.
Both CodeReasoner and CodeBoost only predict inputs and outputs of code without modeling intermediate execution states, rely on pre-collected code reasoning datasets, and treat reasoning as isolated from code generation. 

In this paper, we propose the first work to jointly train execution semantic understanding with code generation using RL, addressing the aforementioned limitations while improving code generation performance.

\begin{figure}[t!] 
    \centering 
    \includegraphics[width=0.83\textwidth]{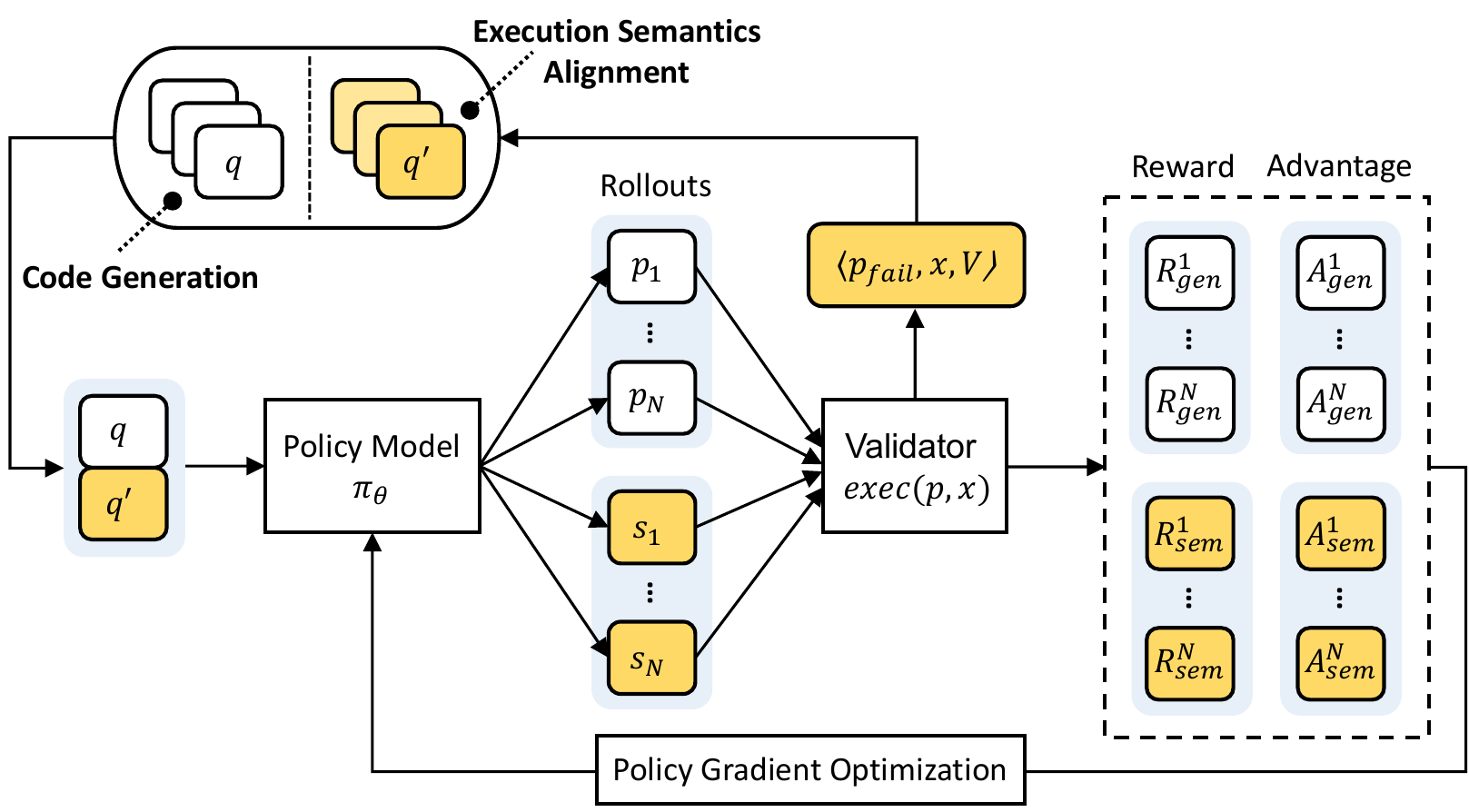} 
    \caption{Overall Pipeline of \ourapproach. Components highlighted in yellow correspond to execution semantics alignment, and the remaining correspond to code generation optimization.} 
    \label{fig:methodology} 
\end{figure}

\section{Methodology}

\subsection{Preliminaries and Definition}

\paragraph{Policy Gradient Optimization.} 

Policy gradient optimization methods are the standard approach for optimizing LLMs within the RLVR framework. Recently, Group Relative Policy Optimization (GRPO)~\citep{DeepSeekMath} has demonstrated exceptional performance in RLVR settings.
Unlike PPO~\citep{PPO}, which requires training an additional value model, GRPO directly estimates advantages through group-normalized rewards, achieving higher computational efficiency.
Specifically, for the programming problem $q$, the model samples $G$ code solutions $\{p_1, p_2, \ldots, p_G\}$. Each solution receives a reward $R_i$ through test case execution (typically binary: 1 for pass, 0 for fail). The GRPO optimization objective is:
\begin{equation}
\begin{aligned}
\mathcal{J}_{\text{GRPO}}(\theta) =&  \mathbb{E}_{q \sim \mathcal{D}, p_i \sim \pi_{\theta}} \Bigg[ \frac{1}{G} \sum_{i=1}^{G} \frac{1}{|p_i|} \sum_{t=1}^{|p_i|} \bigg\{ \min \Big( r_{i,t}(\theta) \cdot \hat{A}_{i,t},\\ 
& \text{clip}(r_{i,t}(\theta), 1-\epsilon, 1+\epsilon) \cdot \hat{A}_{i,t} \Big) - \beta \cdot \mathbb{D}_{\text{KL}}[\pi_\theta \| \pi_{\text{ref}}] \bigg\} \Bigg],
\end{aligned}
\end{equation}
where $r_{i,t}(\theta) = \frac{\pi_\theta(p_{i,t} \mid q, p_{i,<t})}{\pi_{\theta_{\text{old}}}(p_{i,t} \mid q, p_{i,<t})}$ is importance sampling ratio, and $\hat{A}_{i,t} = \frac{R_i - \text{mean}(\{R_1, R_2, \cdots, R_G\})}{\text{std}(\{R_1, R_2, \cdots, R_G\})}$ is the group-normalized advantage estimate.

\paragraph{Execution Semantics}
Execution semantics describes the runtime behavior of a program, \ie, how it processes data, performs computations, and produces results. It provides a foundation for understanding program behavior, debugging and locating errors, optimizing performance, and formally verifying correctness. Formally, a program $p$ can be viewed as a higher-order state transition function $\Phi_p$. This function is composed of atomic transition functions $\phi_t$ corresponding to individual statements in the program. The function takes the current execution state $S_t$ and maps it to the new state $S_{t+1}$ after executing the next instruction.
When this transition function (i.e., the program) operates continuously from an input-determined initial state $S_0$, it generates a sequence of states, \ie, the execution trajectory $\tau$:
\begin{equation}
    \tau = (S_0, S_1, S_2, \dots, S_{\text{final}}),
\end{equation}
where each state $S_t = \{\text{var}_1 \mapsto v_{t,1}, \text{var}_2 \mapsto v_{t,2}, \dots\}$ records the values of all program variables at that step.

We thus formally define the execution semantics of a program under a given input as its complete execution trajectory $\tau$. This trajectory deterministically characterizes the runtime behavior of the program, capturing all intermediate transitions from the initial to the final state.

\subsection{\ourapproach}
Building upon the RLVR framework for code generation, we introduce \ourapproach, which integrates fine-grained execution semantics alignment into the RL training pipeline. The overall workflow of \ourapproach is illustrated in Figure \ref{fig:methodology}.

To integrate execution semantics into the training, we first formalize the concept of execution semantics alignment. This alignment task requires the model to infer the runtime behavior of code, \ie, the execution trajectory $\tau$. However, deriving the complete trajectory is computationally infeasible, as program execution may produce massive intermediate states, particularly within loops where the number of states grows linearly with iterations. Therefore, we propose a tractable approximation:  deriving the final value of each variable as it appears in $\tau$, specifically the value at the variable's last definition point. These final values implicitly encode both the control flow paths taken and the data dependencies resolved during execution, effectively capturing the essential execution semantics while maintaining computational efficiency. Formally, for a program $p$ with variables $V = \{var_1, var_2, \ldots, var_n\}$ and input $x$, we define the execution semantics alignment as:
\begin{equation}
\begin{aligned}
\hat{\mathcal{F}}_p(x) & = \pi_\theta(p, x)\\ 
& \approx \mathcal{F}_p(x) \\
&= \{var_i \mapsto v_{t_i^{\text{last}}, i} \mid var_i \in V\},
\end{aligned}
\end{equation}
where $t_i^{\text{last}} = \max\{t \mid \phi_t \text{ defines } var_i\}$ is the last time step at which variable $var_i$ is defined in the execution trajectory $\tau$, and $\pi_\theta$ denotes the policy model parameterized by $\theta$.

During training, we employ a dual-objective optimization framework
 that simultaneously addresses code generation and execution semantics alignment. Specifically, for each training batch, we construct a mixed prompt distribution $\mathcal{B}_{\text{mixed}} = \alpha \cdot \mathcal{B}_{\text{code}} + (1-\alpha) \cdot \mathcal{B}_{\text{align}}$ by combining code generation prompts and execution semantics alignment prompts with a mixing ratio $\alpha \in [0,1]$.
For each prompt $q_i \in \mathcal{B}_{\text{mixed}}$, the policy model $\pi_\theta$ performs multiple rollouts to generate $N$ samples. These samples represent either complete program solutions $\{p_1, p_2, \ldots, p_N\}$ or execution trace derivation $\{s_1, s_2, \ldots, s_N\}$.

The execution semantics alignment component of $\mathcal{B}_{\text{align}}$ is constructed dynamically from the model's own exploration during training, removing the need for external data and ensuring that the alignment process co-evolves with the model’s code-generation capability.
Specifically, during the rollout phase for code generation, each generated program $p_i$ is executed against the provided test cases to determine its correctness. Failed programs are repurposed for execution semantics alignment training, as they reveal gaps in the model’s understanding of program execution.
For each failed program $p_{\text{fail}}$ from the rollout, we leverage the ground-truth execution semantics $\mathcal{F}_{p_{\text{fail}}}(x)$ obtained during execution on input $x$. We then construct alignment prompts as $q' = \langle p_{\text{fail}}, x, V \rangle$ that challenge the policy model to infer the execution semantics, where the variable names are sequentially specified in the prompt.
Note that the initial training iterations consist entirely of code generation tasks, while subsequent iterations progressively incorporate semantic alignment samples accumulated from failed rollouts.

Following the construction of training batches and rollout generation, we compute rewards for both code generation and execution semantics alignment tasks to guide the model's learning.
For code generation samples, the reward evaluates functional correctness:
\begin{equation}
R_{\text{gen}}^{(i)} = \begin{cases}
1, & \text{if } p_i \text{ passes all test cases}, \\
0, & \text{otherwise}.
\end{cases}
\end{equation}

For execution semantics alignment samples, the reward measures the model's precision in inferring variable states:
\begin{equation}
R_{\text{sem}}^{(i)} = \frac{1}{|V|} \sum_{v_k \in V} \mathbb{1}[\hat{v}_{k}^{\text{final}} = v_{k}^{\text{final},*}],
\end{equation}
where $\hat{v}_{k}^{\text{final}} = \hat{\mathcal{F}}_{p_{\text{fail}}}(x)[v_k]$ is the model's prediction of variable $v_k$'s final value, and $v_{k}^{\text{final},*} = \mathcal{F}_{p_{\text{fail}}}(x)[v_k]$ is the ground-truth value obtained during execution. The indicator function $\mathbb{1}[\cdot]$ returns 1 when prediction matches ground truth, 0 otherwise.

We formulate the final training objective of \ourapproach as a composite function that integrates both code generation and execution semantics alignment:
\begin{equation}
\begin{aligned}
\mathcal{J}_{\text{\ourapproach}}(\theta) = \underbrace{\mathbb{E}_{q \sim \mathcal{B}_{\text{code}}, p \sim \pi_\theta} \left[ r(\theta) \cdot A_{\text{gen}} \right]}_{\text{Code Generation Optimization}} + \underbrace{\mathbb{E}_{q' \sim \mathcal{B}_{\text{align}}, \hat{\mathcal{F}}_{p_{\text{fail}}}(x) \sim \pi_\theta} \left[ r'(\theta) \cdot A_{\text{sem}} \right]}_{\text{Execution Semantics Alignment}},
\end{aligned}
\end{equation}
where $r(\theta)$ and $r'(\theta)$ are the importance sampling ratios for code generation and execution semantics alignment, respectively. The advantages $A_{\text{gen}}$ and $A_{\text{sem}}$ are computed using group normalization based on their corresponding rewards $R_{\text{gen}}^{(i)}$ and $R_{\text{sem}}^{(i)}$ within their respective groups, following the GRPO framework.

\ourapproach establishes a learning framework grounded in the formal execution semantics: code generation learns to synthesize the state transition function $\Phi_p$ while execution semantics alignment learns to understand $\Phi_p$. Synthesizing $\Phi_p$ entails generating code that realizes the desired state transformations, whereas understanding $\Phi_p$ through inferring the trajectory $\tau$ reveals how these transformations evolve program state during execution. Through joint optimization, \ourapproach transcends learning from surface-level code patterns, instead fostering a deeper understanding of the bidirectional relationship between code structure and its execution dynamics.

\section{Experiments}
We present extensive experiments spanning three code-related tasks, five representative datasets, three different LLMs, and three different RL algorithms to demonstrate the effectiveness of our approach. Furthermore, we conduct comprehensive analyses from four perspectives, including training dynamics, ablation studies, probing analysis, and case studies (presented in Appendix \ref{appendix:cases}), to provide deeper insights of \ourapproach.

\subsection{Experiment Setup}

\paragraph{Training Details.}
We use prime code data as our training dataset~\citep{PRIME_Zero}, sourced from APPS~\citep{APPS}, CodeContests~\citep{alphacode}, TACO~\citep{TACO}, and Codeforces~\citep{Codeforces}, comprising 27K coding problems along with their corresponding test cases.
By default, we employ Qwen2.5-Coder-7B-Instruct~\citep{hui2024qwen2} as the base model throughout our experiments. For the implementation of the RL algorithm, we leverage the VeRL framework~\citep{sheng2024hybridflow}. The training configuration includes a batch size of 128, a mini-batch size of 64, a learning rate of 1e-06, and a maximum of 1000 training steps. For each problem, we generate 8 rollout samples with a maximum response length of 8192 tokens. All experiments are conducted on a cluster of 8 NVIDIA A100 80G GPUs.
Regarding the hyperparameters for \ourapproach, we incorporate execution semantics alignment prompts at a ratio of 0.4 per batch. To ensure fair comparison, all other RL algorithms are configured with the same parameter settings as those used in \ourapproach.

\paragraph{Evaluation Details.}  
Consistent with prior work~\citep{li2025codeprm,codereasoner,wang2025codeboost}, we evaluate on three standard code generation benchmarks: HumanEval~\citep{chen2021evaluating}, LeetCode~\citep{leetcode}, and LiveCodeBench~\citep{jain2024livecodebench}, using pass@1 as the evaluation metric.
To further examine whether execution-based semantic alignment benefits other code-related tasks, we also evaluate on Code Reasoning and Test Output Generation. For Code Reasoning, we use the LiveCodeBench-Reason~\citep{jain2024livecodebench} benchmark, which requires models to generate function outputs given Python functions and inputs, with accuracy as the evaluation metric. For Test Output Generation, we use LiveCodeBench-Test~\citep{jain2024livecodebench} benchmark, where models must generate outputs based on problem descriptions and inputs, which is a particularly challenging test generation task, also evaluated using accuracy.
All evaluations use greedy sampling with temperature set to 0.0.

\paragraph{Baselines.}  
In addition to the base model and standard GRPO method~\citep{DeepSeekMath}, we compare \ourapproach against two categories of methods, all trained upon the same base model.
The first category comprises four recently proposed post-training models and methods for code generation, including:
1) \textbf{OlympicCoder}~\citep{openr1} is fine-tuned using chain-of-thought traces distilled from DeepSeek-R1 on competitive programming problems.
2) \textbf{OCR-Qwen-7B}~\citep{ocrqwen} is another open-source code model distilled from DeepSeek-R1, trained on an extensive dataset of up to 730,000 samples with reasoning trajectories.
3) \textbf{Skywork-OR1}~\citep{skywork} is a code generation model trained via large-scale RLVR following the DeepSeek-R1 pipeline.
4) \textbf{CodePRM}~\citep{li2025codeprm} leverages process reward models in RL for code generation.
Since code reasoning involves execution semantics inference, we also compare against three state-of-the-art code reasoning methods, including: 5) \textbf{CODEI/O}~\citep{licodeio}, 6) \textbf{CodeReasoner}~\citep{codereasoner}, 7) \textbf{CodeBoost}~\citep{wang2025codeboost}.
\begin{table*}[t!]
\centering
\caption{Performance of \ourapproach compared to baselines. \textbf{Bold} indicates the best result, and \underline{underline} indicates the second-best result for each metric.}
\resizebox{\textwidth}{!}
{
\begin{tabular}{lcccccc}
\toprule
\multirow{2}{*}{Approach}
 & \multicolumn{4}{c}{Code Generation} & Code Reasoning & Test Output Generation \\ \cmidrule(r){2-5} \cmidrule(r){6-6} \cmidrule(r){7-7}
 & HumanEval & LeetCode & LiveCodeBench & Average & LiveCodeBench-Reason & LiveCodeBench-Test \\
\midrule
Qwen2.5-Coder-7B-Instruct & \underline{88.4} & 50.6 & 34.3 & 57.8 & 60.8 & 48.8 \\
GRPO & 87.2 & \underline{60.0} & \underline{35.4} & \underline{60.9} & 66.0 & 48.4 \\
\hdashline
\multicolumn{7}{l}{\textbf{Code Generation Baselines}}\\
OlympicCoder & 75.6 & 45.3 & 30.9 & 50.6 & 68.5 & 31.1 \\
OCR-Qwen-7B-Instruct & 86.8 & 53.3 & 33.0 & 57.7 & 44.1 & 28.3 \\
Skywork-OR1 & 87.2 & \underline{60.0} & 33.8 & 60.3 & 69.5 & 48.2  \\
CodePRM & \underline{88.4} & 52.8 & 34.8 & 58.7 & 62.4 & 48.1 \\
\hdashline
\multicolumn{7}{l}{\textbf{Code Reasoning Baselines}} \\
CODEI/O & 86.0 & 41.7 & 27.2 & 51.6 & 57.2 & 41.3 \\
CodeReasoner & \underline{88.4} & 50.0 & 34.8 & 57.7 & \underline{78.5} & \textbf{65.1} \\
CodeBoost & 87.2 & 53.3 & 34.6 & 58.4 & 67.2 & 52.0 \\
\hdashline
\ourapproach & \textbf{90.9} & \textbf{63.3} & \textbf{36.9} & \textbf{63.7} & \textbf{85.0} & \underline{53.2} \\
\bottomrule
\end{tabular}} \label{tab:main_results}
\end{table*}

\subsection{Experiment Results}

\paragraph{Performance of \ourapproachbf.}
Table \ref{tab:main_results} presents the main results of \ourapproach compared to baselines. Our approach achieves SOTA performance on all code generation benchmarks, consistently outperforming recently proposed post-training methods for code generation. Our method also demonstrates strong generalization to code-related tasks, achieving the best performance on Code Reasoning and second-best on Test Output Generation. We observe that RL-based methods mostly outperform SFT-based methods, \ie, GRPO, Skywork-OR1, and CodePRM surpass OlympicCoder and OCR-Qwen-7B, while CodeReasoner and CodeBoost outperform CODEI/O. This trend underscores the advantage of RL for both in-domain and out-of-domain tasks. While code reasoning-oriented methods tend to underperform on code generation, GRPO—commonly applied to code generation—yields only limited improvement in reasoning ability, our approach successfully bridges this gap by combining code generation training with execution semantics alignment, achieving the best results on both code generation and reasoning tasks. Moreover, CodeReasoner achieves the best performance on Test Output Generation. We analyze that this is due to its additional pre-RL training phase that leverages extensive data distilled from powerful teacher models, enhancing its capability in this specific task.

\begin{table*}[t!]
\centering
\caption{Performance of \ourapproach on different series and size LLMs.}
\resizebox{\textwidth}{!}
{
\begin{tabular}{lcccccc}
\toprule
\multirow{2}{*}{Approach} & \multicolumn{4}{c}{Code Generation} & Code Reasoning & Test Output Generation \\ \cmidrule(r){2-5} \cmidrule(r){6-6} \cmidrule(r){7-7}
 & HumanEval & LeetCode & LiveCodeBench & Average & LiveCodeBench-Reason & LiveCodeBench-Test \\
\midrule
\textbf{LLaMA-3.1-8B-Instruct} & 68.9 & 12.8 & 10.9 & 30.9 & 40.7 & 27.6 \\
GRPO & 59.8 & 21.1 & 11.9 & 30.9 & 26.9 & 25.6 \\
\ourapproach & \textbf{70.7} & \textbf{34.4} & \textbf{21.1} & \textbf{42.1} & \textbf{40.9} & \textbf{27.7} \\
\hdashline
\textbf{Qwen2.5-Coder-7B-Instruct} & 88.4 & 50.6 & 34.3 & 57.8 & 60.8 & 48.8\\
GRPO & 87.2 & 60.0 & 35.4 & 60.9 & 66.0 & 48.4 \\
\ourapproach &\textbf{90.9} & \textbf{63.3} & \textbf{36.9} & \textbf{63.7} & \textbf{85.0} & \textbf{53.2}\\
\hdashline
\textbf{Qwen2.5-Coder-1.5B} &70.1  &17.8  & 12.5 & 33.5 & 31.1 & 33.1 \\
GRPO & 65.2 & 17.8 & 17.4 & 33.5 & 28.0 & 30.0 \\
\ourapproach & \textbf{75.0} & \textbf{37.8} & \textbf{17.4} & \textbf{43.4} & \textbf{34.9} & \textbf{34.0} \\
\bottomrule
\end{tabular}}\label{tab:diff_LLMs}
\end{table*}

\paragraph{Application on Various LLMs.}
To demonstrate the generalizability of \ourapproach, we apply it to different LLMs, including LLaMA-3.1-8B-Instruct~\citep{llama31}, Qwen2.5-Coder-7B-Instruct~\citep{hui2024qwen2}, and Qwen2.5-Coder-1.5B~\citep{hui2024qwen2}. As shown in Table~\ref{tab:diff_LLMs}, our method consistently outperforms the standard GRPO baseline across all benchmarks and model variants. Notably, while GRPO sometimes struggles with training stability (e.g., showing performance degradation on LLaMA-3.1-8B), our execution-based approach achieves robust improvements across different model families and sizes. On LLaMA-3.1-8B-Instruct, \ourapproach achieves an average absolute improvement of 11.2\% over the GRPO baseline in code-generation performance.

\begin{table*}[t!]
\centering
\caption{Performance of \ourapproach on RL Algorithms.}
\resizebox{\textwidth}{!}
{
\begin{tabular}{lcccccc}
\toprule
\multirow{2}{*}{Approach} & \multicolumn{4}{c}{Code Generation} & Code Reasoning & Test Output Generation \\ \cmidrule(r){2-5} \cmidrule(r){6-6} \cmidrule(r){7-7}
 & HumanEval & LeetCode & LiveCodeBench & Average & LiveCodeBench-Reason & LiveCodeBench-Test \\
 \midrule
GRPO & 87.2 & 60.0 & 35.4 & 60.9 & 66.0 & 48.4 \\
+ \ourapproach & \textbf{90.9} & \textbf{63.3} & \textbf{36.9} & \textbf{63.7} & \textbf{85.0} & \textbf{53.2} \\
\hdashline
PPO & 88.4 & 45.0 & 29.6 & 54.3 & 61.0 & 39.5 \\
+ \ourapproach & \textbf{89.6} & \textbf{61.1} & \textbf{34.5} & \textbf{61.7} & \textbf{78.5} & \textbf{52.7} \\
\hdashline
REINFORCE++ & 82.3 & 53.9 & 32.5 & 56.2 & 58.7 & 47.2 \\
+ \ourapproach  & \textbf{92.1} & \textbf{63.9} & \textbf{33.8} & \textbf{63.3} & \textbf{78.9} & \textbf{51.1} \\
\bottomrule
\end{tabular}}\label{tab:diff_RL}
\end{table*}

\paragraph{\ourapproachbf with Other RL Algorithms.}
Our approach can be seamlessly integrated with various RL algorithms. We evaluate its effectiveness when combined with GRPO~\citep{DeepSeekMath}, PPO~\citep{ppo_meta}, and REINFORCE++~\citep{rein2024gpqa}. As shown in Table \ref{tab:diff_RL}, our approach consistently enhances all three RL algorithms across all benchmarks. 
Our method delivers the most substantial improvement to PPO (+7.4\% average on code generation), even surpassing the gains achieved on GRPO (+2.8\%).

\subsection{Analysis}

\begin{wrapfigure}{r}{0.49\textwidth}
    \vspace{-0.3cm}
    \centering 
    \includegraphics[width=0.48\textwidth]{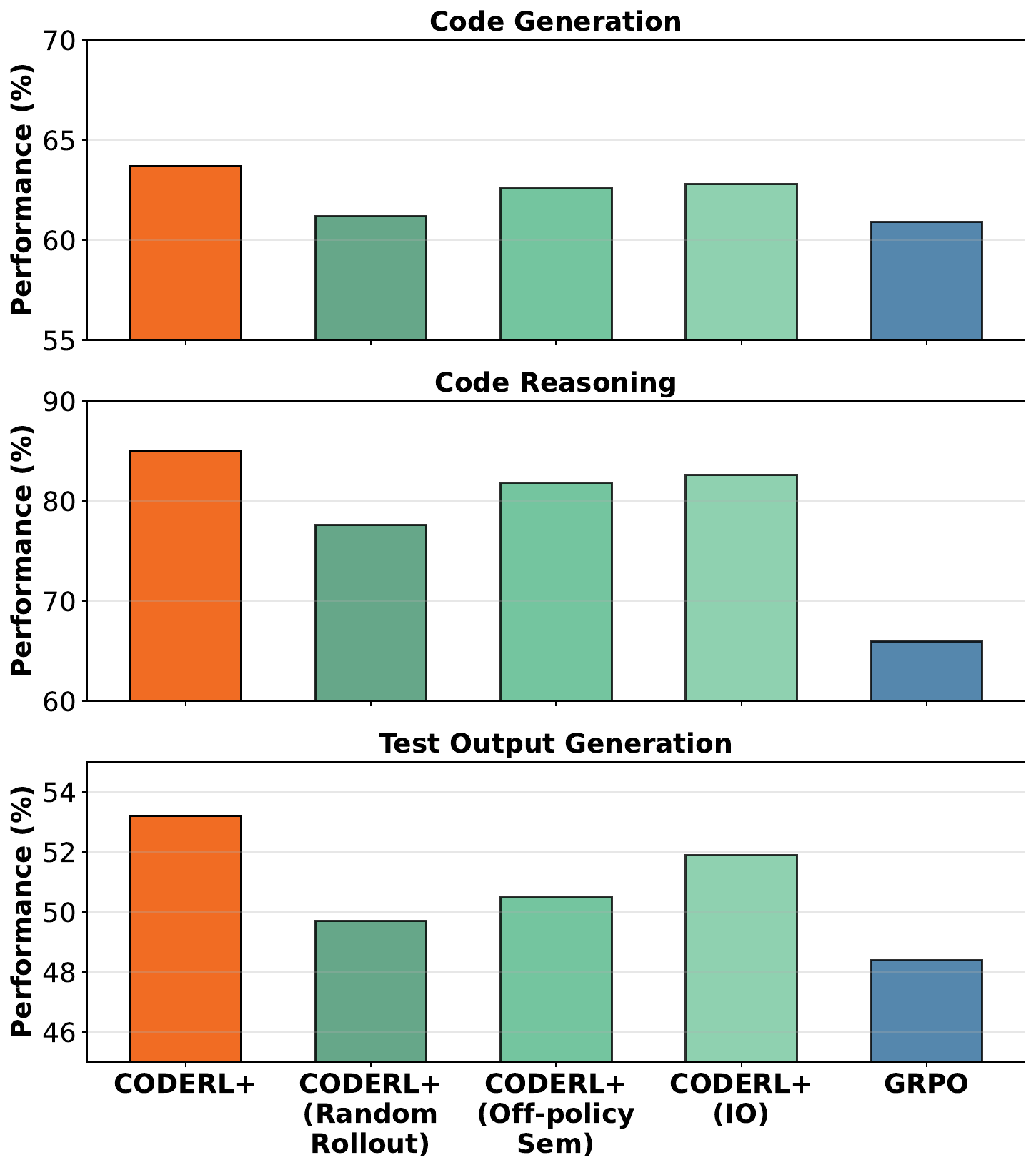} 
    \caption{Result of Ablation Study.} 
    \label{fig:ablation_study} 
\end{wrapfigure}

\paragraph{Ablation Study.}
We conduct ablation studies to validate three key design choices in our approach, with results shown in Figure \ref{fig:ablation_study}. First, to verify the effectiveness of leveraging failed rollout codes, we compare against \ourapproach (Random Rollout), which does not distinguish between correct and incorrect samples. The performance drop demonstrates that selectively using failed rollouts provides more informative learning signals. We observe that \ourapproach (Random Rollout) achieves high execution semantics alignment rewards during training, indicating these samples lack sufficient challenge to drive meaningful improvements. Second, we evaluate the importance of on-policy execution semantics alignment by comparing with \ourapproach (off-policy Sem), which pre-constructs alignment examples before training, using code generation training data. The superior performance of our on-policy approach confirms that execution semantics alignment that evolves with model training is more effective. Finally, \ourapproach (IO), which only supervises input-output pairs rather than fine-grained variable trajectories, shows degraded performance across all tasks, highlighting the value of dense supervision signals from intermediate execution states. These ablations collectively demonstrate that each component of our approach contributes meaningfully to its overall effectiveness.

\begin{figure*}[t!] 
    \centering 
    \includegraphics[width=0.98\textwidth]{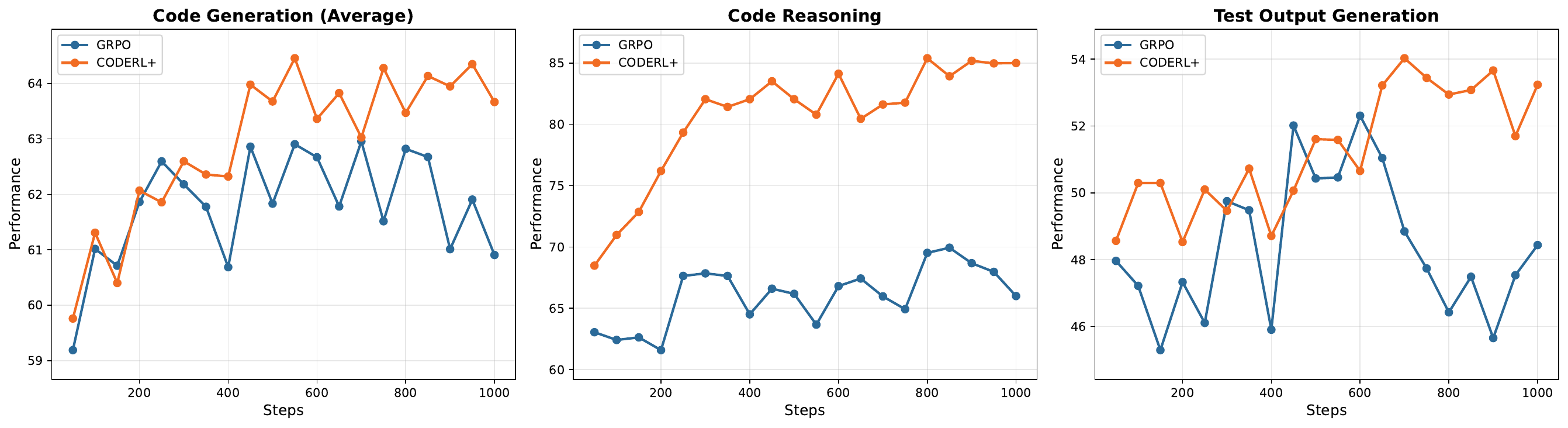} 
    \caption{Training dynamics of \ourapproach and GRPO.} 
    \label{fig:training_dynamics} 
\end{figure*}

\paragraph{Training Dynamics.}
Figure \ref{fig:training_dynamics} illustrates the training dynamics of our method and baseline GRPO across three tasks. The results demonstrate that our approach consistently outperforms GRPO under the same number of training steps (and thus equal training data), with the performance gap widening in later stages. The widening gap could be explained by GRPO training solely on code-generation objectives, without explicitly modeling execution semantics. Beyond code generation, our method exhibits a substantial advantage over GRPO on code reasoning tasks. Notably, for test output generation task, GRPO exhibits minimal improvement throughout training, reflecting the large domain gap between this task and the training distribution. In contrast, our approach demonstrates steady improvement on this challenging task, benefiting from its enhanced understanding of execution semantics acquired through execution semantic alignment during training.

\begin{figure}[t!] 
    \centering 
    \includegraphics[width=0.9\textwidth]{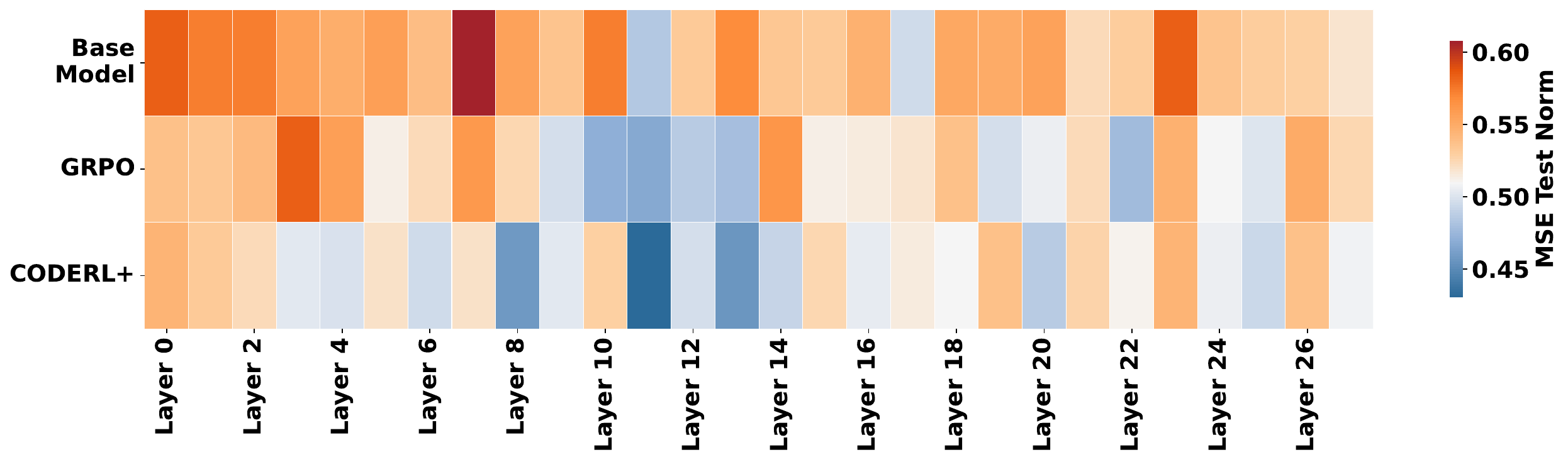} 
    \caption{Probing results of \ourapproach on HumanEval code generation benchmark.} 
    \label{fig:probing} 
\end{figure}

\paragraph{Probe Analyses.}
To investigate the impact of incorporating execution semantics alignment on the model's internal representations, we conduct a probe experiment. Probes are supervised models trained to predict specific properties from learned representations~\citep{hewitt2019designing,lee2021predicting}. In our case, we analyze whether the model's representations implicitly encode execution semantics, specifically, whether the representations of variables in generated code can predict their runtime values.
Our experimental setup (detailed in Appendix \ref{probe_setup}) employs linear regression probes to predict intermediate variable values from the hidden states extracted at variable token positions. We evaluate three models: the base model, GRPO-trained model, and our \ourapproach-trained model. The probe performance, measured by Mean Squared Error (MSE)~\citep{mse} on normalized variable values, serves as a quantitative indicator of how well the model's internal representations align with actual execution semantics.

The experimental results shown in Figure \ref{fig:probing} demonstrate that \ourapproach achieves lower MSE across all model layers compared to both the base model and GRPO, indicating stronger alignment between textual representations and execution semantics. This improvement is particularly pronounced in the middle layers, where semantic understanding is typically encoded. These findings provide empirical evidence that our execution semantics alignment mechanism effectively guides the model to develop internal representations that better capture the execution behavior of code, rather than merely learning textual patterns.

\section{Conclusion}
In this work, we presented \ourapproach, which addresses the fundamental semantic gap between how LLMs learn code (through textual patterns) and how code actually works (through execution semantics). By incorporating execution semantics alignment into RLVR training, our method moves beyond sparse pass/fail rewards to provide direct learning signals that explicitly connect code's textual form with its execution behavior. Extensive experiments demonstrate that \ourapproach delivers substantial improvements across multiple code generation benchmarks and generalizes effectively across different code-related tasks, LLMs, and RLVR algorithms.

\section*{Acknowledgments}
This research is supported by the National Natural Science Foundation of China under Grant No. 62192733, 62192730, 62192731, the National Key R\&D Program under Grant No. 2023YFB4503801, and the Beijing Major Science and Technology Project under Contract No. Z251100008425005.

\bibliography{preprint}
\bibliographystyle{preprint}

\newpage
\onecolumn
\appendix

\section{Case Study}
\label{appendix:cases}

\begin{figure*}[h!] 
    \centering \includegraphics[width=\textwidth]{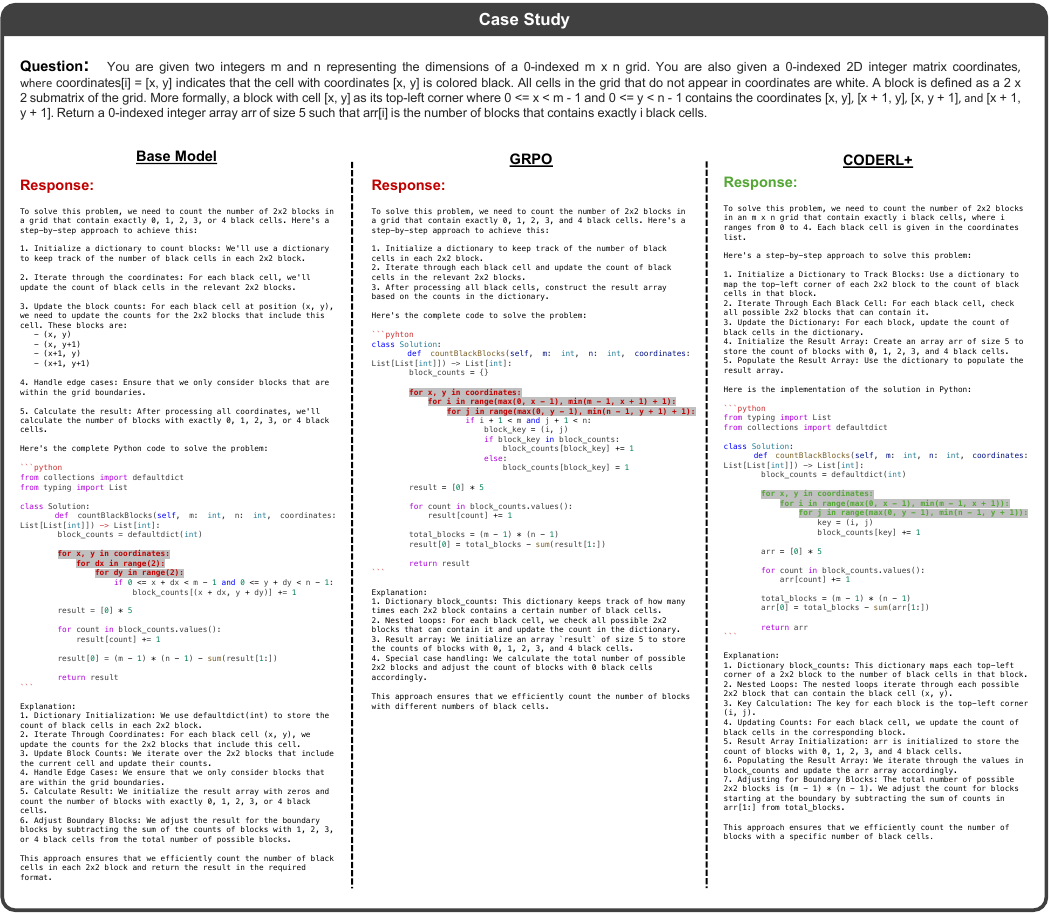} 
    \caption{Cases of base model (Qwen2.5-Coder-7B-Instruct), GRPO, and \ourapproach.} 
    \label{fig:case_study} 
\end{figure*}

We conduct a case study that demonstrates our approach can mitigate logical errors in code generation. Figure~\ref{fig:case_study} presents the number-of-black-blocks problem, where different methods exhibit distinct enumeration strategies. The base model incorrectly iterates with ranges that miss left and top blocks while over-counting right and bottom ones. GRPO partially improves but still uses incorrect ranges for $i \in \text{range}(\max(0, x - 1), \min(m - 1, x + 1) + 1)$, leading to overcounting. In contrast, our proposed \ourapproach correctly identifies that for each black cell at position $(x, y)$, it can belong to at most 4 possible $2 \times 2$ blocks, and properly implements the enumeration logic by checking the valid range boundaries, ensuring accurate counting without missing or double-counting any blocks. This demonstrates \ourapproach's effectiveness in learning precise logical patterns for code generation tasks.

\section{Setup of Probe Analyses}\label{probe_setup}

This experiment is based on the HumanEval code generation dataset and investigates three models: Base Model (Qwen2.5-Coder-7B-Instruct), GRPO, and \ourapproach. We first process the original dataset by decomposing each task, which contains multiple test cases, into several "single-example" tasks where each prompt includes only one input-output example. Subsequently, for each original task, we randomly split all its corresponding sub-tasks into a training set and a test set for the probe, using an 8:2 ratio.

Our experimental pipeline follows a "generate-execute-extract" paradigm. First, we prompt the three models to generate Python code for each single-example prompt. Next, we execute the generated code with its corresponding input example, capturing and recording the final values of all numerical intermediate variables within the function upon its completion. Finally, we feed the generated code back into its respective model, identify the token corresponding to the last occurrence of each traced variable, and extract the hidden state vector for that token from all model layers to serve as input features for the probe.

We train an independent linear regression model as a probe for each layer of each model. Given that the intermediate variables can vary significantly, we preprocess the data by normalizing the target variable values on a per-problem basis using Min-Max scaling to the range [-1, 1]. The normalization parameters are computed exclusively from the training set. All probes are trained for 10 epochs using the Adam optimizer with a learning rate of 1e-3, minimizing the Mean Squared Error (MSE) loss. For evaluation, our metric is the MSE in the normalized space on the test set. A lower MSE value indicates a better alignment between the model's internal representations and the code's execution semantics.

\section{Experiment Setup of Execution Trace Inference Task.}

We evaluate on the LiveCodeBench Code Reasoning task with an extended dataset. While the original task requires predicting a function's return value given its input and code, we extend it to also predict the final value of each intermediate variable at the end of its lifetime. This extension enables us to assess the model's understanding of code execution traces.
Specifically, we leverage the existing inputs and code from the LiveCodeBench Code Reasoning task, execute the code, and extract intermediate variables along with their final values at the end of their respective lifetimes.
The evaluated models include: Base Model (Qwen2.5-Coder-7B-Instruct), GRPO, and \ourapproach.
We use Exact@1 as the evaluation metric, which measures the proportion of cases where all intermediate variables and the final function return value are correctly predicted.
The evaluation prompt is as follows, where variables in blue are to be replaced with actual content:

\begin{promptbox}{Execution Trace Inference Task}
Given the following Python Code and Input, predict:

1) The code's output value (final\_output).

2) The final values of the listed local variables at the moment the code outputs.
\\\\
Python code: \verb|```python|\textcolor{blue}{\{code\}}\verb|```|

Input: \textcolor{blue}{\{test\_input\}}

Target local variables: \textcolor{blue}{\{local\_variable\_name\}}
\\\\
\textit{Instructions:}

1. First, write a reasoning section explaining step-by-step how the code executes with the given Input.

2. Do not include any JSON in the reasoning section.

3. On the LAST line only, output a strict JSON object with the required format. Example of final answer (LAST line only):
\{"final\_output": 3, "variables": \{"cnt": 2, "buf": [1, 2]\}\}
\end{promptbox}

\section{Implementation Details of Execution Semantics Alignment}
In implementation, the variable set $V$ is restricted to variables of primitive types (\eg, integers, floats, and strings), whose values can be deterministically represented and compared for reward computation. The ground-truth variable states $\mathcal{F}_{p_{\text{fail}}}(x)$ are obtained by executing the program and capturing the final values of target variables upon termination\footnote{The variable state extraction can be integrated into the existing execution process used for computing code generation rewards, requiring only a single execution per program and thus introducing minimal overhead.}. Furthermore, programs that encounter runtime errors are filtered out, as they do not yield valid execution trajectories. Only programs with semantic errors, \ie, code that executes normally but produces incorrect outputs, are selected for alignment. Such semantic errors represent the majority of code generation errors in LLMs.

\section{Additional Benchmark Results}\label{appendix:additional_benchmarks}

To further evaluate the generalizability of \ourapproach, we conduct additional experiments on two more code generation benchmarks: MBPP~\citep{mbpp} and LiveBench~\citep{livebench}. As shown in Table~\ref{tab:additional_benchmarks}, our method consistently outperforms the base model and GRPO baseline on both datasets in terms of pass@1, demonstrating that \ourapproach generalizes well beyond the primary evaluation benchmarks.

\begin{table}[h]
\centering
\caption{Performance of \ourapproach on additional code generation benchmarks.}
\begin{tabular}{lcc}
\toprule
Approach & MBPP & LiveBench \\
\midrule
Qwen2.5-Coder-7B-Instruct & 70.7 & 46.4 \\
GRPO & 72.1 & 46.9 \\
\ourapproach & \textbf{76.2} & \textbf{50.6} \\
\bottomrule
\end{tabular}\label{tab:additional_benchmarks}
\end{table}

\section{Limitations}
Our work has three limitations. First, computational constraints limited our evaluation to models up to 8B parameters, which may affect the generalizability of our conclusions to larger-scale LLMs. Second, we did not perform hyperparameter tuning due to the high computational cost of RL post-training, as each training run requires roughly three days. We followed prior work for training-related hyperparameters and empirically set the single hyperparameter specific to our method, maintaining this configuration across all experiments where it consistently yielded improvements. Third, while the execution semantics alignment component incurs additional computational overhead compared to standard RL algorithms, our experiments demonstrate that our approach achieves superior performance within comparable computational budgets (measured by training steps) and reaches higher performance ceilings as training progresses.

\end{document}